# An Investigation of the Incidences of Repetitive Strain Injury among computer Users in Nigeria

Olabiyisi Olatunde[1], Akingboye Yusuff[2], Abayomi-Alli Adebayo[3], Izilien Fred[4], Adeleke Iyiola[5]

[1] Computer Science and Engineering Department, Ladoke Akintola University of Technology, Ogbomoso, Oyo State Nigeria.
[2,4,5] Electrical and Computer Engineering Department, Igbinedion University Okada, Edo State, Nigeria.
[3] Computer Science Department, Federal University of Agriculture, Ogun State, Nigeria.

**Abstract**

Computer has been incorporated into day to day activities of almost every field of human endeavour, offices to different shops. Therefore many people are now working with computer for longer hours of time. There is no doubt that this incorporation of computer has helped users a lot but it also brings problems to the users. One of the problems is Repetitive Strain Injury (RSI).

Five hundred and thirty one (531) questionnaires were personally administered to different categories of people that use computer in various works of life, ranging from banking sector, civil service, educational sector, health sector to private sector. The distribution cut across different professions. A statistical analysis was conducted on the data obtained using frequency distribution, Pearson Correlation and Linear Regression.

The result obtained showed that 94.3% of the respondents suffered pain from one or more parts of the body. 86.8% of the respondents suffered from eyestrain, 63.9% suffered from low back pain, 67.4% with wrist pain, 64.7% finger pain while the least suffered pain was foot pain which only 19% responded positively to it. There are significant relationships between duration of computer usage, type of chair used, type and size of monitor used and the incidence of RSI. RSI modeled was formulated through linear regression which showed that a unit change in computer will result in corresponding 1.76 unit increases in RSI and a unit change in ergonomic deficiency will also result in corresponding 0.66 increases in RSI.

The existence of RSI was established and it was discovered that the more time spent on the computer system, the more the proximity of having strain or pain in one or more part(s) of the body.

## 1. Introduction

In recent times, computer has become a common tool that is being used by almost every individual from various field of human endeavour. This is due to the fact that computer offer a lot of different services and facilities to help the users to perform and complete the tasks more efficient and effective. Injuries due to the usage of computer system has been recognized worldwide and several actions that involved repetitive or forceful movements and the maintenance of constrained or awkward postures have also been associated with musculoskeletal disorder (Repetitive Strain Injury). Poor posture, prolong starring at computer screen, repetitive reaching for mouse, sitting position and type of chairs has been discovered to affect the lower back, eyes, arm of computer users [1].

Repetitive strain injury (RSI), also called cumulative trauma disorder (CTD), occupational overuse syndrome, or work related upper limb disorder (WRULD), as conditions resulting from continuous use of a tool, e.g. Computer, guitar, knife, etc. over a long period of time or activity that requires repeated movements. It is a syndrome that affects muscles, tendons and nerves in the hands, arms and upper back. The medically accepted condition in which it occurs is when muscles in these areas are kept tense for very long periods of time, due to poor posture and/or repetitive motions as agreed by various researchers [1] [2][3].

Over the years, computer-related injuries (known as repetitive strain injuries or RSIs) have increasingly plagued the modern office workplace, debilitating hundreds of thousands of workers, causing pain, impairment and, in some cases, disability [4]. Several studies linked overuse of computer system with the increase of Repetitive strain injury (Herbert, 2005). With the increase in computer dependency in various works of life, individual who spend more period of time on the computer are prone to a greater risk of developing RSI and it has been observed that individual suffering from RSI tend to be less productive.

RSI is a national tragedy because it has drastically disrupted the lives and livelihoods of many people all over the world [5]. This tragedy is compounded by the fact that it is preventable in most of the cases. However, we need to convince computer users in Nigeria that RSI is a serious public health problem. There is need for proper information on the extent of Repetitive Strain Injury among computer users in Nigeria and the various ways in which it can be prevented in other to aid policy making. In this context, effort has been made to find the root causes of RSI among computer users in Nigeria and provide appropriate awareness to the career threatening syndrome.

### 1.1 Related Works

Nokubonga Slindele in his study on awkward working postures and precision performance as an example of the relationship between ergonomics and production quality stated that posture adopted has a direct influence on performance variables which are associated with productivity and the quality of product output. Therefore, it provides evidence that ergonomics interventions are relevant and critical in contributing positively to organizational goals simultaneously to taking cognizance of worker needs. And stated further that the results he obtained indicated that, though, precision tasks are considered to be 'light' tasks, the static muscular contractions required to stabilize individuals in awkward postures are high enough to be a cause for concern with regards to the early onset of fatigue and the precipitation of musculoskeletal injuries. This emphasizes the importance and necessity of workstation design to simultaneously consider performance outcomes and the strain on workers [6].

Adedoyin et al did a survey of computer users across six federal university campuses in Nigeria. A frequency analysis was done and his result showed that Low back pain and neck pain were found to be the highest pain complaint with 74% and 73% respectively. 67% of the respondents complained of wrist pain, followed by finger pain (65%), shoulder pain (63%) and general body pain (61%). The knee and foot pains were the least complaints reported with 26% and 25% respectively. In terms of pain severity, low back pain, finger pain, neck pain and shoulder pain are rated to be moderate, while all other joints were said to be of mild pain. His study indicated that low back pain, neck pain and upper limbs are the common disorders complaints among the users. He attributed the cause of the pains to bad ergonomics among the users [7].

Allen E. Akhowa did a survey of computer users in University of Benin, Nigeria on the issue of Occupational Overuse Syndrome (OOS). The data for his study was collected by means of a structured questionnaire administered to respondents (computer users) at the University of Benin campuses comprising staff and students. His results showed that low back pain, neck pain, headache, shoulder pain and eyestrain, are the most prevalent OOS symptoms/pains. He recommended that there is need for computer workplaces to improve on their designs towards finding a lasting solution to the hazardous problem [3].

Sanusi B. also did a survey of incidence of RSI on the students of University of Ibadan, his findings showed that there is correlation and significant relationship between postures maintained by computer users and the incidences of Repetitive Strain Injury and concluded that not keeping a good posture while working on the computer is a major cause of RSI [8].

Many of the studies conducted in Nigeria to investigate the occurrence of RSI focused on its occurrence within the University community but this study survey general computer users from various works of life.

## 2. Research Methodology

Social survey design was used for this research in other to obtain relevant information from respondents on the incidents of Repetitive Strain Injury (RSI) among computer users in Nigeria. Questionnaire was the instrument used for data collection in this study.

The questionnaire was designed and self-administered to different categories of computer users in various works of life in Nigeria, ranging from banking sector, civil service, educational sector, health sector to private sector. The distribution was ensured to cut across different field of professionalism. The questionnaire designed was divided into four sections, namely: Bio-data, general question, ergonomic factors, Body strain and ergonomics exercise. This was structured with a checklist of responses.

The data obtained from the questionnaires administered were analyzed using the Statistical Package for the Social Sciences (SPSS) software version 16.0 for windows. The analysis was done in three facets: descriptive analysis through the frequencies procedure which produced frequency tables that displayed both the numbers and percentages of cases for each observed value of variables.

Hypothesis were tested using Pearson Correlation to establish whether or not there is relationship between incidence of RSI and duration of computer usage (years of computer usage and hours of computer usage at a stretch), type of chair used, size and types of monitor used.

### 2.1 Research Hypothesis

Hypothesis 1
$H_0$: There is no significant relationship between the duration of computer usage and the incidence of Repetitive Strain Injury.

$H_A$: There is significant relationship between the duration of computer usage and the incidence of Repetitive Strain Injury.

Hypothesis 2
$H_0$: There is no significant relationship between the type of chair used when working on the computer and the incidence of Repetitive Strain Injury.

HA: There is significant relationship between the type of chair used when working on the computer and the incidence of Repetitive Strain Injury.

Hypothesis 3
$H_0$: There is no significant relationship between the size and type of monitor used and the incidence of eye strain.

$H_A$: There is significant relationship between the size and type of monitor used and the incidence of eye strain.

2.1.1 Criteria for Rejection and Acceptance of Hypothesis

$\alpha = 0.01$
If $p <= 0.01$ the null hypothesis is rejected
If $p > 0.01$ null hypothesis is accepted
Where $\alpha$ is the level of significant and p-value is the probability that the observed correlation coefficient r was seen by chance.

Linear Regression analysis was done to establish the rate at which ergonomic risk factors explains the incidence of RSI.

## 2.2 Linear Regression Model Specification

The following models which aided in seeing the causal relationship between the dependent and independent variables are stated below:
RSIIndex = f (EF)
Where RSIindex – Index of Repetitive Strain Injury
EF – Ergonomic Factors
$\alpha_1$: coefficient of the independent variables

## 3. Results

### 3.1 Demographic Information of Respondents

Five hundred and thirty one questionnaires were administered. The demographic distribution of the respondents according to table 1 indicated the age and gender distribution of the respondents. 59.9% of the respondents were males while 40.1% were females. 1.9% of the respondents were between the age distribution of

The functional form of the model stated above is shown below:
RSIIndex = $\alpha + \alpha_1$EF

A positive relationship is expected between EF and RSIindex because various literature revealed that poor ergonomics has great impact on Repetitive strain Injury. 1-17 years, 47.8% between 18-30 years, and 43.3% between 31-45 years while 7% were 46 years and above.

### 3.2 Type of Chair Used

Table 2 revealed that 63.7% of the respondents usually sit on fixed chair while working on the computer and the remaining 36.3% sits on adjustable chair throughout their daily computational period.

Table 1: Demographic Information of Respondents

|  | Frequency | Percent | Valid Percent | Cumulative Percent |
|---|---|---|---|---|
| **Age** |  |  |  |  |
| 1-17 years | 10 | 1.9 | 1.9 | 1.9 |
| 18-30 years | 254 | 47.8 | 47.8 | 49.7 |
| 31-45 years | 230 | 43.3 | 43.3 | 93.0 |
| 46-65 years | 37 | 7.0 | 7.0 | 100.0 |
|  | 531 | 100.0 | 100.0 |  |
| **Gender** |  |  |  |  |
| MALE | 318 | 59.9 | 59.9 | 59.9 |
| FEMALE | 213 | 40.1 | 40.1 | 100 |
| Total | 531 | 100 | 100 |  |

Table 2: Type of User's Chair

|  |  | Frequency | Percent | Valid Percent | Cumulative Percent |
|---|---|---|---|---|---|
| Valid | Fixed | 338 | 63.7 | 63.7 | 63.7 |
|  | Adjustable | 193 | 36.3 | 36.3 | 100.0 |
|  | Total | 531 | 100.0 | 100.0 |  |

### 3.3 Duration of Computer Usage

Table 3 showed that respondents who used their computer system for 3-4 hours at a stretch takes the highest percentage of 37.5% followed by those that use it for more than 6 hours which is 0.2% greater than respondents using computer for 4-6 hours with 28.6%. Only 3.8% use computer for 1-2 hours while 1.3% of the respondents rarely use computer. The table also showed that 78.9% of the respondents have been using the computer system for more than 4 years

Table 3: Duration of Computer Usage

|  |  | Frequency | Percent | Valid Percent | Cumulative Percent |
|---|---|---|---|---|---|
| | **Years of Computer Usage** | | | | |
| **Valid** | <1 year | 9 | 1.7 | 1.7 | 1.7 |
| | 1-3 years | 103 | 19.4 | 19.4 | 21.1 |
| | 4-7 years | 239 | 45.0 | 45.0 | 66.1 |
| | >7 years | 180 | 33.9 | 33.9 | 100 |
| | Total | 531 | 100 | 100 | |
| | **Hours of Computer Usage at stretch** | | | | |
| **Valid** | Rarely | 7 | 1.3 | 1.3 | 1.3 |
| | 1-2 hours | 20 | 3.8 | 3.8 | 5.1 |
| | 3-4 hours | 199 | 37.5 | 37.5 | 42.6 |
| | 4-6 hours | 152 | 28.6 | 28.6 | 71.2 |
| | >6 hours | 153 | 28.8 | 28.8 | 100.0 |
| | Total | 531 | 100.0 | 100.0 | |

Respondents were asked to specify the pains experienced and their location. Table 4 above showed the percentage of people that experience different types of pains on parts of the body.

### 3.4 Pain Incidences

People suffering from eyestrain has the highest percentage with 83.8% followed by Low back pain and wrist pain respectively. Foot pain has the least percentage of people suffering from it followed by ankle pain and elbow pain respectively.

Table 4: Pain Incidences Reported by Respondents

| | Number of Respondents = 531 | | | | |
|---|---|---|---|---|---|
| **Pains Suffered During Computer use** | Never(0) | Mild(1) | Moderate(2) | Severe(3) | **Total Affected |
| **Lowback Pain** | 30.70% | 34.50% | 28.80% | 6.00% | 69.30% |
| **Neck Pain** | 32.80% | 43.30% | 20.70% | 3.20% | 67.20% |
| **Shoulder Pain** | 55.90% | 30.30% | 11.30% | 2.40% | 44.00% |
| **Elbow Pain** | 75.30% | 17.90% | 5.80% | 0.90% | 24.60% |
| **Wrist Pain** | 32.60% | 31.80% | 31.10% | 4.50% | 67.40% |
| **Finger Pain** | 35.40% | 23.20% | 36.20% | 5.30% | 64.70% |
| **Headache** | 54.80% | 28.10% | 15.80% | 1.30% | 45.20% |
| **Arm Pain** | 40.30% | 21.80% | 34.80% | 3.00% | 59.60% |
| **Body Pain** | 46.00% | 31.30% | 21.10% | 1.70% | 54.10% |
| **Hip Pain** | 68.00% | 17.30% | 13.90% | 0.80% | 32.00% |
| **Ankle Pain** | 81.00% | 9.80% | 8.70% | 0.60% | 19.10% |
| **Knee Pain** | 61.60% | 17.10% | 19.40% | 1.90% | 38.40% |
| **Foot Pain** | 81.00% | 9.40% | 9.00% | 0.60% | 19.00% |
| **Eye Strain** | 16.20% | 10.70% | 57.10% | 16.00% | 83.80% |
| | **never excluded | | | | |

## 3.5 Tests for Correlation
### 3.5.1 Duration of computer Usage and Incidence of RSI

There is a positive and strong correlation between the number of hours spent on the computer at a stretch and the incidence of repetitive strain injury (r=0.611, n=531). The result showed a significant correlation (p=0.007, p<0.01) between the number of hours spent on the computer at a stretch and the incidence of repetitive strain injury. The null hypothesis is rejected and alternate hypothesis accepted.

There is also a positive and strong correlation between the number of years spent on the computer and the incidence of repetitive strain injury (r=0.581, n=531). The result showed a significant correlation (p=0.001, p<0.01) between the number of years spent on the computer and the incidence of repetitive strain injury. The null hypothesis is rejected and alternate hypothesis accepted (see table 5).

Table 5: Correlations of Duration of Computer Usage and Pain Incidence

|  |  | Pain Incidence on Parts of the Body during and after Computer Use | Computer Use Hour at a Stretch | Years of Computer Usage |
|---|---|---|---|---|
| Pain Incidence on Parts of the Body during and after Computer Use | Pearson Correlation | 1 |  |  |
|  | Sig. (2-tailed) |  |  |  |
|  | N | 531 |  |  |
| Computer Use Hour at a Stretch | Pearson Correlation | .611** | 1 |  |
|  | Sig. (2-tailed) | .007 |  |  |
|  | N | 531 | 531 |  |
| Years of Computer Usage | Pearson Correlation | .581** | .607** | 1 |
|  | Sig. (2-tailed) | .001 | .000 |  |
|  | N | 531 | 531 | 531 |

**. Correlation is significant at the 0.01 level (2-tailed).

### 3.5.2 Chairs Used and Incidence of RSI

The types of chairs used on the computer system was negatively correlated with incidence of low back pain(r=.538, n=532, p=.003 P<.01). The correlation is significant at 0.01 levels. This indicate that the more use of fixed chairs was associated with incidence of repetitive strain injury. The null hypothesis is therefore rejected and the alternate hypothesis accepted (see table 6)

Table 6: Correlations of Types of user's chair and incidence of low back pain

|  |  | Incidence of Low back Pain | Type of User's Chair |
|---|---|---|---|
| Incidence of Low back Pain | Pearson Correlation | 1 | .538** |
|  | Sig. (2-tailed) |  | .003 |
|  | N | 531 | 531 |
| Type of User's Chair | Pearson Correlation | .538** | 1 |
|  | Sig. (2-tailed) | .003 |  |
|  | N | 531 | 531 |

**. Correlation is significant at the 0.01 level (2-tailed).

### 3.5.3 Type and Size of Monitor Used and Incidence of RSI

There is a positive correlation (r=0.442) between the types of monitor used and incidence of Eye Strain. The result is significant at p=0.004 (p<.01) therefore the null hypothesis is rejected and alternate hypothesis accepted (see table 7). The result showed that there is a significant correlation between monitor screen size and the incidence of repetitive strain injury, r= -.519, n=452, p<.01, two tails. Monitor screen size is positively associated with the incidence of Repetitive Eye Strain. The null hypothesis is therefore rejected and the alternate hypothesis accepted (see table7)

Table 7: Correlations of Incidence of Eyestrain, Types of Monitor Used and Monitor Screen Size

|  |  | Incidence of Eye Strain | Types of Monitor Used | User's Monitor Screen Size |
|---|---|---|---|---|
| Incidence of Eye Strain | Pearson Correlation | 1 | .442** | -.519** |
|  | Sig. (2-tailed) |  | .004 | .001 |
|  | N | 531 | 452 | 452 |
| Types of Monitor Used | Pearson Correlation | .442** | 1 | .381** |
|  | Sig. (2-tailed) | .004 |  | .009 |
|  | N | 452 | 452 | 449 |
| User's Monitor Screen Size | Pearson Correlation | -.519** | .381** | 1 |
|  | Sig. (2-tailed) | .001 | .009 |  |
|  | N | 452 | 449 | 452 |

### 3.6 Regression Model to Establish Effect of Ergonomic Factors on RSI

RSI index (0 - 1) that is (no strain - strained)
*Relation*
Dependent: RSI index
Independent: Ergo risk Factors
*Analysis*
R=0.808

The level of correlation among RSIindex (dependent variables) and ErgoFactors (predictor) is 0.907. This result indicates that there is a strong correlation between Ergofactors and RSIindex based on RSIindex at 90.7%. $R^2 = 0.822$.

This showed that the analysis is adjudged accurate at 82.2% and that about 82.2% variation in the dependent variable is being explained by the variation in the independent variables while the remaining 17.2% variation cannot be explained by the variation in the independent variable. (Good).

#### 3.6.1 Test for Regression Parameters
$H_0: \alpha_1 = \alpha_2$; $H_1: \alpha_1 \neq \alpha_2$
$F_{cal} = \frac{MS Regression}{MS Error(residual)}$

$F_{cal} = \frac{2.093}{0.042} = 69.453$
$F_{cal} = 69.453$

Test statistics F- Test
$F_{tab} = F_{(V-1)\ (N-V)}^{\alpha}$
$F_{(2)(3)}^{0.05} = P_{value} = 0.000^a$

Decision Rule
If $F_{cal} < F_{tab}$, accept $H_0$ (accept the null hypothesis)
If $F_{cal} > F_{tab}$, do not accept $H_0$ (accept alternative hypothesis)

Comparism
Fcal = 69.453
Ftab = 0.000

Conclusion
While $F_{cal}$ (69.453) > $F_{tab}$ (0.000), we do not accept $H_0$, therefore we accept $H_1$ and conclude that the regression parameter is significant for RSIindex using ErgoFactors.

Model
RSIindex = -2.67 + 0.664ErgoFactors

Interpretation

1. Sign (+ or -)
   The model shows a positive relation between ErgoFactors and RSI index. This implies that ErgoFactor is a corresponding measure of RSI index, that is, increases in non-compliance of ergonomic way of using the computer system and setting up workstation, the more the RSI.
2. Change
   A unit change in ErgoFactor will result in correspondent 66.4% increase in RSI. In addition the model has high RSI index based ErgoFactor. This is true because non-compliance with the correct use of computer system, correct way of setting up workstation result in RSI.

## 4. Discussion

It was found in this study that many people suffered from pains from one part of the body to another during and after working on the computer system, some suffered pain from only one part of the body while majority suffered pains from more than one parts of the body. 95.4% of the respondents said they have incidence of pains while only 5.5% claimed they do not suffered from any form of pain.

The result also showed that more than 80% of the population of computer users in Nigeria lies between the age of 18 and 45 years. Over 63% of the population under study used a fixed chair while working on the computer system while the rest sit on ergonomically designed chairs that are adjustable. This indicates that more people will have pains related to bad sitting position while working for a period of time at a stretch. It was also discovered from the survey that 87.4% of the respondents who suffered from one pain or the other are those that usually work on the computer system for more than three hours at a stretch. This indicates that the more hours one spent on the computer system, the higher the proximity of having RSI from different parts of the body. The same percentage of respondents suffers joint pain while typing on the computer system for up to 4 hours.

## 5. Conclusion

Our findings have generally linked long duration of computer usage, usage of ergonomically defective chairs, and bad ergonomic work station to incidence of Repetitive Strain injury. We also discovered that with increased computer literacy in the developing countries, many users' workplaces still lack proper design, management strategies, and task design. This showed that proper attention has not been given to the need to minimizes or eradicate workplace hazard.

4.1 Recommendations

1. Enlightenment programs should be created especially in the university communities in Nigeria to educate students on preventive measures to overcome repetitive strain injury.
2. Establishment of ergonomic programs such as seminars and trainings, these bring awareness and consciousness on the safety measures for the working conditions to all computer users to remedy and prevent workplace injuries.
3. Computer users should cultivate the habit of taking short fixed breaks during their daily computer work to perform different types of ergonomic exercises.
4. The media houses should also create regular awareness on this career threatening syndrome.
5. Workplaces should put proper preventive guidelines in place while those already having ergonomics guidelines should monitor for continual research adherent to it.

Therefore, our findings generally recommend the need to improve computer workplaces in terms of design, management strategies, creation of awareness program and usage of ergonomically designed equipment. When this is done, it will go a long way to avoiding poor posture during computer work and thereby reduce the incidence of RSI.

## Biography

**Olabiyisi, S. O**. received his B. Tech., M. Tech and Ph.D degrees in Mathematics from LadokeAkintola University of Technology, Ogbomoso, Nigeria, in 1999, 2002 and 2006 respectively. He also received M.Sc. degree in Computer Science from University of Ibadan, Ibadan, Nigeria in 2003. He is currently an Associate Professor in the Department of Computer Science and Engineering, LadokeAkintola University of Technology, Ogbomoso, Nigeria. He has published in reputable journals and learned conferences. DrOlabiyisi is a full member of Computer Professional (Registration) Council of Nigeria (CPN). His research interests are in computational mathematics, theoretical computer science, information systems and performance modelling and simulation.

**Akingboye, A.Y**. graduated from the Ladoke Akintola University of Technology, Ogbomoso with a Master of Technology (M.Tech) and Bachelor of Technology (B.Tech) degrees in Computer Science in 2012 and 2005 respectively. His research interests include Microprocessor Systems and Human Computer Interaction. He is presently with the Department of Electrical and Computer Engineering at Igbinedion University Okada.

**Abayomi-Alli, A.** obtained his B.Tech Degree in Computer Engineering from Ladoke Akintola University of Technology (LAUTECH), Ogbomoso in 2005, MSc Computer Science from the University of Ibadan, Nigeria in the 2009. He started his career at Igbinedion University Okada, Nigeria as a Graduate Assistant in 2007 before moving to the Federal University of Agriculture Abeokuta (FUNAAB), Nigeria in 2011. His current research interests include microprocessor systems and applications, biometrics, image quality assessment and machine learning.

**Izilein, F. A.** has a B.Eng and M.Eng Degree in ElectricalElectronics. He presently lectures at the Department ofElectrical and Computer Engineering in Igbinedion University Okada.

**Adeleke, I. A**. graduated from the Ladoke Akintola University of Technology, Ogbomoso with a Bachelor of Technology (B.Tech) degrees in Computer Engineering in 2005. His research interests include Microprocessor Systems and Electronics. He is presently with the Department of Electrical and Computer Engineering at Igbinedion University Okada.